\documentclass[aps,pra,superscriptaddress,twocolumn]{revtex4}

\usepackage{amsmath}
\usepackage{amssymb}
\usepackage{graphicx}
\usepackage{dcolumn}

\begin{document}

\title{Atomic Quantum Corrals for Bose-Einstein Condensates}
\author{Hongwei Xiong}
\email{xionghongwei@wipm.ac.cn}
\affiliation{State Key Laboratory of Magnetic Resonance and Atomic and Molecular Physics,
Wuhan Institute of Physics and Mathematics, Chinese Academy of Sciences,
Wuhan 430071, China}
\affiliation{Center for Cold Atom Physics, Chinese Academy of Sciences, Wuhan 430071, P.
R. China}
\author{Biao Wu}
\email{bwu@aphy.iphy.ac.cn}
\affiliation{Institute of Physics, Chinese Academy of Sciences, Beijing 100190, China}
\date{\today }

\begin{abstract}
We consider the dynamics of Bose-Einstein condensates in a corral-like
potential. Compared to the electronic quantum corrals, the atomic quantum
corrals have the advantage of allowing direct and convenient observation of
the wave dynamics. Our numerical study shows that this advantage not only
allows people to explore the rich dynamical structures in the density
distribution but also makes the corrals useful in many other aspects. In
particular, the corrals for atoms can be arranged into a stadium shape for
the experimental visualization of quantum chaos, which has been elusive with
the electronic quantum corrals.
\end{abstract}

\maketitle





\section{introduction}

Quantum corrals were first demonstrated by arranging iron adatoms into a
ring on a copper surface \cite{eleccroo,Heller_rmp}. Several years later, an
optical analogy to the electronic quantum corrals was proposed theoretically
and realized in experiment by a skillful arrangement of nanoscale pillars
\cite{theo-opt,exp-opt}. However, in both of the quantum corrals only static
wave properties were studied experimentally. Since the density image was
obtained by sequential scanning over a period of time, the experimental
study of wave dynamics in these corrals is difficult if not impossible.

In this paper, we study the dynamics of atomic quantum corrals for
Bose-Einstein condensates (BECs). Since the density distribution of a BEC
can be imaged snapshot by snapshot with a charge coupled device (CCD), the
atomic quantum corrals offer a great advantage over their electronic and
optical counterparts, namely, the possible experimental study of the wave
dynamics inside the corrals. Our numerical simulation shows that rich
dynamical structures of a condensate can arise in the quantum corrals due to
reflection and interference. Moreover, with the great deal of experimental
control over the condensates \cite{Ketterle-Rev,Rev}, one is possible to
study novel quantum behaviors, which are unimaginable for electronic or
optical quantum corrals. For example, the dynamic evolution of a quantized
vortex confined inside quantum corrals can be studied. 

Of particular importance, the atomic quantum corrals proposed here can be
used as a laboratory to study quantum chaos. Even since the first creation
of quantum corrals, it has been pursued to build stadium-shaped quantum
corrals to visualize experimentally the \textquotedblleft scar" states, a
signature of quantum chaos \cite{Heller_nature}. The stadium-shaped quantum
corrals were built; however, the visualization of quantum chaos has remained
elusive because the quantum corrals are too \textquotedblleft leaky" for
electrons \cite{Heller_nature}. Our atomic quantum corrals are very flexible
and can be made to minimize the \textquotedblleft leakage\textquotedblright\
so that be used to explore quantum chaos experimentally. As an example, we
demonstrate with numerical simulation that with atomic quantum corrals of
stadium shape, one should be able to visualize experimentally the quantum
chaotic behavior predicted in Ref. \cite{stadium}. We also show how the
interference between two BECs can be destroyed by chaos.

The paper is organized as follows. In Sec. II, we consider the wave packet
dynamics for a BEC in a corral-like potential. In particular, the vortex
evolution in a corral-like potential is studied. In Sec. III, we study the
dynamic quantum chaos for a BEC in stadium-shaped quantum corrals. A brief
summary and discussion is given in the last section.

\begin{figure}[tbp]
\includegraphics[width=0.9\linewidth,angle=0]{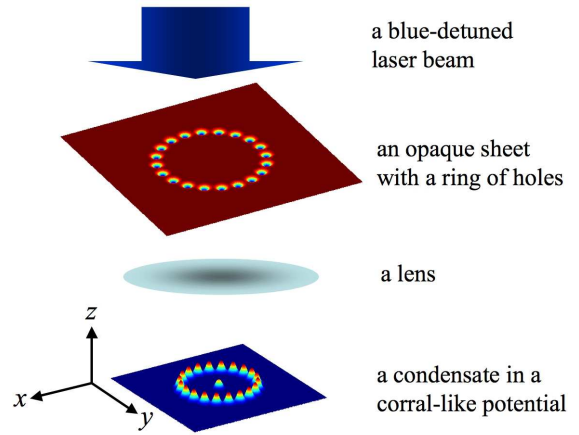}
\caption{Scheme to build quantum corrals for a Bose-Einstein
condensate (see text). }
\end{figure}

\section{wave packet dynamics in corral-like potential}

We consider here a scheme to create quantum corrals for a BEC illustrated in
Fig. 1. An opaque sheet with a ring of holes is placed between a lens and a
blue-detuned laser beam to create a corral-like repulsive potential for the
condensate. The lens is to focus and adjust the sizes of the
\textquotedblleft corrals". The holes in the opaque sheet are identical with
a radius around or larger than 50$\mu $m, which is large enough so that the
diffraction of the laser beam can be ignored. This setup is similar to the
one used to create a rotating quasi-2D optical lattice by a mask with a set
of holes \cite{Tung} and a random potential \cite{random} for BECs. One may
also create this kind of quantum corrals by spatial light modulator or
microlens array\cite{Lv}.

To simulate the dynamics of the cigar-shaped BEC in the corrals, we can
integrate out the axial degree of freedom. In this case, the BEC dynamics is
described by a two dimensional Gross-Pitaevskii (GP) equation \cite{Rev}:%
\begin{eqnarray}
i\frac{\partial \Phi }{\partial t} &=&-\left( \frac{\partial ^{2}}{\partial
x^{2}}+\frac{\partial ^{2}}{\partial y^{2}}\right) \Phi +  \notag \\
&&V_{oc}\Phi +V_{ht}\Phi +g_{2D}\left\vert \Phi \right\vert ^{2}\Phi \,,
\label{GPequation}
\end{eqnarray}%
where $\Phi $ is a normalized wave function. The above equation has been
made dimensionless with the length unit $L_{0}$ and the energy unit $%
E_{0}=\hbar ^{2}/2mL_{0}^{2}$. The time unit is then $T_{0}=\hslash /E_{0}$.
$V_{oc}$ is the corral-like potential while $V_{ht}=m\omega _{\perp
}^{2}L_{0}^{2}\left( x^{2}+y^{2}\right) /2E_{0}$ with $\omega _{\perp }$
being the harmonic frequency in $x$ and $y$ directions. The dimensionless
coupling constant $g_{2D}=2\sqrt{2\pi }N\hbar ^{2}a/ml_{z}E_{0}L_{0}^{2}$,
where $a$ is the $s-$wave scattering length and $l_{z}=\sqrt{\hbar /m\omega
_{z}}$ with $\omega _{z}$ being the trapping frequency in the $z$ direction.
We numerically solve this GP equation to illustrate the wave dynamics inside
the quantum corrals. In the numerical calculations, we consider $2\times
10^{4}$ \textrm{Na} atoms in the condensate and use $L_{0}=5$ $\mathrm{\mu m}
$, $T_{0}=1.8\mathrm{ms}$, and $\omega _{z}=4\times 2\pi $ \textrm{Hz}.

The $i$th Gaussian potential $V_{i}\left( x,y,z\right) $ created by a
focused laser beam propagating along $z$ direction has the following form%
\begin{equation}
V_{i}\left( x,y,z\right) \sim \frac{P}{\pi \sigma ^{2}\left( z\right) }e^{-%
\left[ \left( x-x_{j}\right) ^{2}+\left( y-y_{j}\right) ^{2}\right] /\sigma
^{2}},
\end{equation}%
where $P$ is the total power of a laser beam. $\sigma \left( z\right)
=\sigma _{0}\sqrt{1+z^{2}/z_{R}^{2}}$ with the Rayleigh length $z_{R}=2\pi
\sigma _{0}^{2}/\lambda $. In the following calculations, $\sigma _{0}=5$ $%
\mathrm{\mu m}$. For a focused laser beam with wave length $\lambda =500$
\textrm{nm}, we have $z_{R}=314$ $\mathrm{\mu m}$, which is much larger than
the length (about $10$ $\mathrm{\mu m}$ for the parameters in this paper) of
the cigar-shaped condensate. In this situation, along the $z$ direction, the
condensate feels almost the same corral-like potential in the $x-y$ plane.
Thus, the two dimensional GP equation can be used very well to study the
dynamics of a condensate in the corral-like potential.

To show clearly the fundamental properties of the atomic quantum corrals, we
calculate numerically the evolution of the condensate confined in
circle-shaped quantum corrals, which can be described as a series of
Gaussian potentials $V_{oc}=\sum\limits_{j=1}^{M}\gamma e^{-\left(
(x-x_{j})^{2}+(y-y_{j})^{2}\right) /\sigma ^{2}}$ with $\{x_{j},y_{j}\}$
distributed uniformly along a circle of radius $R$. With the parameters $%
\gamma =20$, $M=20$, $R=10$, $\sigma =1$, and $\omega _{\perp }=70\times
2\pi $, we first get the ground-state wave function by using the widely used
ITP (imaginary time propagation) method. Then the evolution of the
condensate is solved numerically from the GP equation (\ref{GPequation}),
after suddenly switching off the harmonic trap. In this situation, the
condensate will evolve under the confinement of the corral-like potential.
In Fig. 2a, both the initial density distribution $\left\vert \Phi
\right\vert ^{2}$ and the corral-like potential are shown. In Figs. 2b-2f,
the evolution of $\left\vert \Phi \right\vert ^{2}$ is given, after suddenly
switching off the harmonic trap. The rich and colorful structures in $%
\left\vert \Phi \right\vert ^{2}$ originate from the following two physical
mechanisms:

\begin{figure}[tbp]
\includegraphics[width=0.75\linewidth,angle=270]{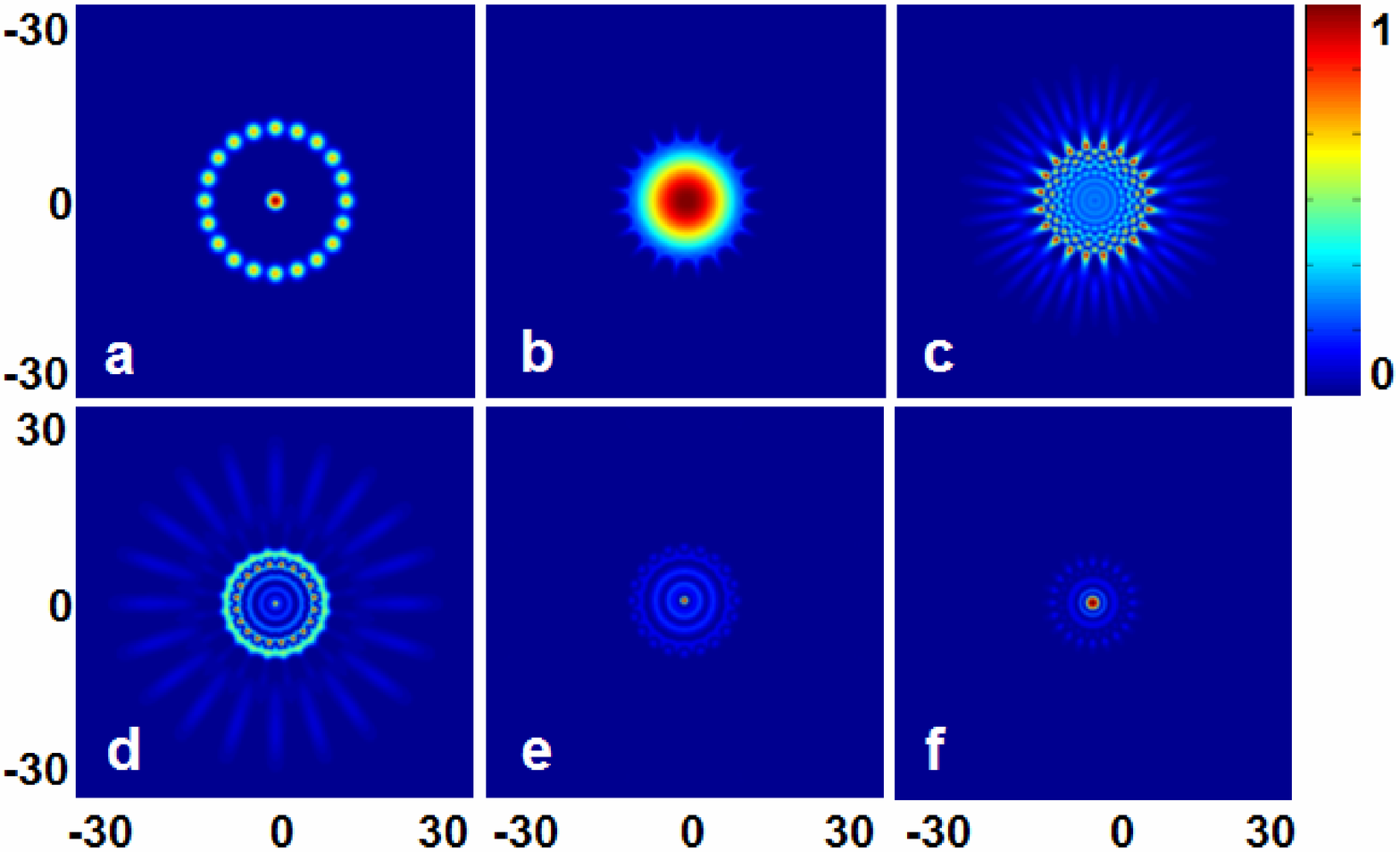}
\caption{Density distribution of a condensate in a corral-like
potential at different times: (a) $t=0$, (b) $t=1$, (c) $t=2$, (d)
$t=3$, (e) $t=4$, (f) $t=5$. The coordinates are in unit of $L_{0}$.
In Fig. a, the corral-like potential is also shown.}
\end{figure}

(i) After switching off the harmonic trap, and with the expansion of the
condensate, the condensate will be reflected by the corral-like potential.
The expanding and reflected condensates will overlap, and lead to clear
interference patterns. In Figs. 2c-2f, a series of ring-shaped interference
fringes are clearly shown.

(ii) In Figs. 2c and 2d, the density distribution shows sunflower-like
structure. This sunflower-like structure is due to the discrete
characteristic of the corral-like potential. For the expanding condensate,
the corral-like potential can be regarded as $M$ scattering sources,
arranged along a circle. This discrete characteristic gives important
modulation on the interference fringes. Because of this, the ring-shaped
interference fringe is composed of a series of small peaks. In Figs. 2c and
2d, the number of small peaks distributed along an interference fringe is
found to be exactly $20$. By varying $M$, we have verified numerically that
the number of small peaks distributed along a ring-shaped interference
fringe is always equal to $M$, the number of corrals.

The coherent interaction between atoms and external field (such as laser and
magnetic field) can provide us important opportunity to study various
dynamic processes \cite{Ketterle-Rev}. We consider here the evolution of the
condensate after the sudden decreasing of the harmonic frequency to $\omega
_{\perp }=17.5\pi $. The evolution of the density distribution is shown in
Fig. 3. The density distribution in Figs. 3d-3f shows typical flower-like
structure in the kirigami. Because of the confinement of the harmonic trap,
we see a quasi-periodic oscillation behavior. In Fig. 3i, the standard
deviation $\delta $ of the density distribution is given, which shows
further this quasi-periodic behavior. Without the corral-like potential and
interatomic interactions, the time period would be exactly $\pi /\left(
\omega _{\perp }T_{0}\right) $. From Fig. 3i, the time period is slightly
smaller than $\pi /\left( \omega _{\perp }T_{0}\right) $ because of the
corral-like potential and interatomic interactions.

\begin{figure}[tbp]
\includegraphics[width=0.75\linewidth,angle=270]{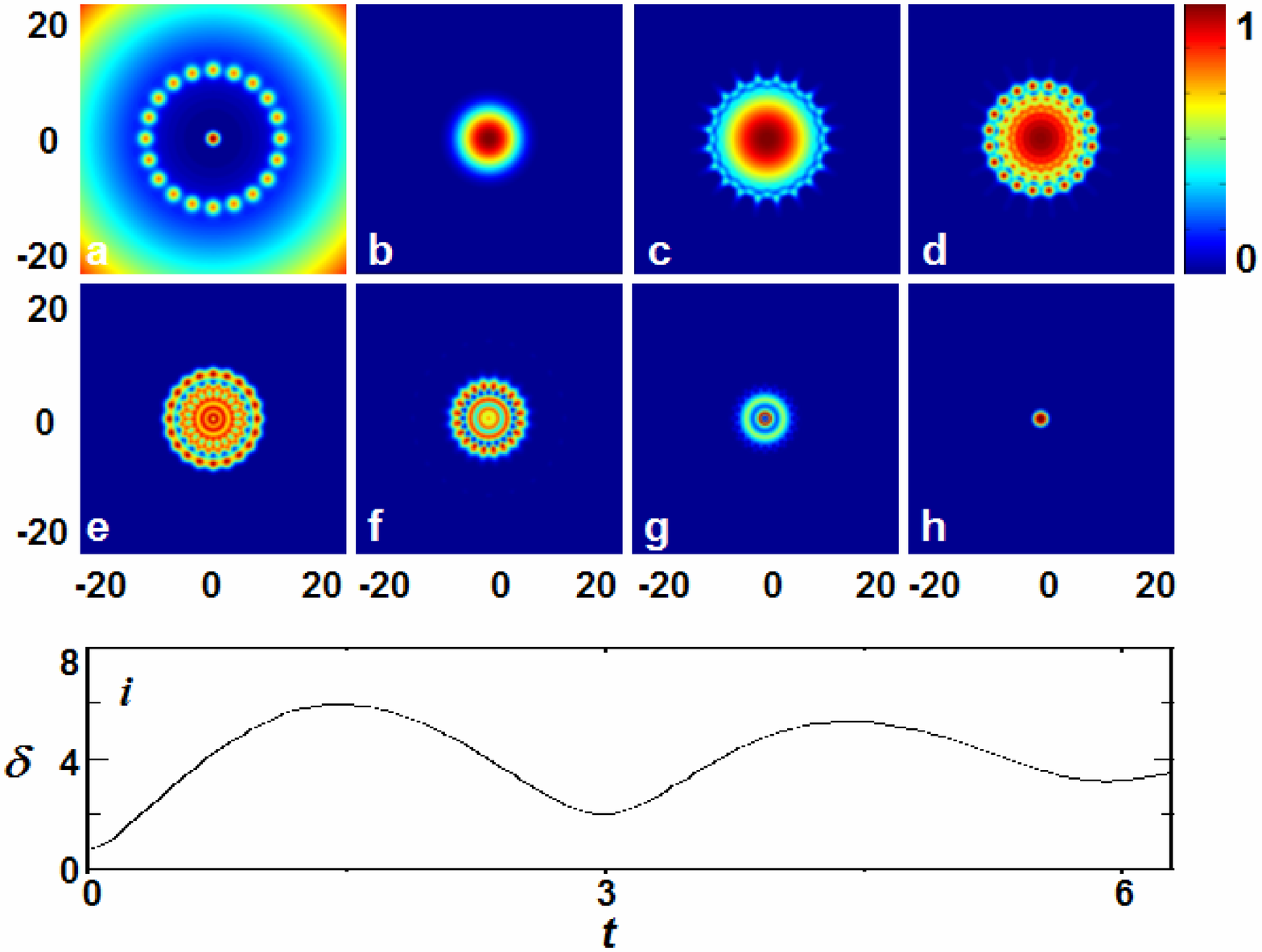}
\caption{Evolution of the density distribution in the presence of
both a corral-like potential and a harmonic potential. The dynamic
evolution is due to the sudden decreasing of the harmonic frequency.
(a) $t=0$, (b) $t=\protect\pi /5$, (c) $t=2\protect\pi /5$, (d) $t=\protect%
\pi /2$, (e) $t=3\protect\pi /5$, (f) $t=7\protect\pi /10$, (g) $4\protect%
\pi /5$, (h) $\protect\pi $. Here the coordinates are in unit of $L_{0}$.
Fig. i gives the time evolution of the standard deviation $\protect\delta $
of the density distribution. In Fig. a, the corral-like potential and
harmonic potential are also shown. The height of these two potentials has
been adjusted in Fig. a.}
\end{figure}

The corral-like potential plays a key role in the interference fringes of
the density distribution. Without the corral-like potential, the decreasing
of the harmonic frequency will only lead to a width oscillation of the
condensate \cite{Rev}. We have verified this by numerical calculations,
which show that there are no interference fringes for this case.

We now consider the evolution of a quantized vortex in the circle-shaped
quantum corrals. In our calculations, we use $\gamma =20$, $M=20$, $R=10$, $%
\sigma =1$, and $\omega _{\perp }=70\times 2\pi $. The initial vortex state
in the presence of both the corral-like and harmonic trapping potentials is
obtained numerically from the ITP method with a trial wave function, which
has the general form $f\left( \sqrt{x^{2}+y^{2}}\right) e^{\pm i\theta }$
\cite{Rev}. The signs `$+$' and `$-$' in $e^{\pm i\theta }$ denote the
rotational directions of the vortex.

The evolution of the vortex in the corrals is shown in Fig. 4. The evolution
is obtained numerically from the GP equation (\ref{GPequation}). After
switching off the harmonic trap, and with the expansion of the condensate,
the condensate is reflected by the corral-like potential. Similarly to the
preceding results, the expanding and reflected condensates will overlap, and
lead to clear interference patterns. These interference patterns are clearly
seen in Figs. 4a-4f, where Figs. 4a-4c show the evolution for the case of $%
e^{i\theta }$ while Figs. 4d-4f give the evolution for the case of $%
e^{-i\theta }$.

\begin{figure}[tbp]
\includegraphics[width=0.75\linewidth,angle=0]{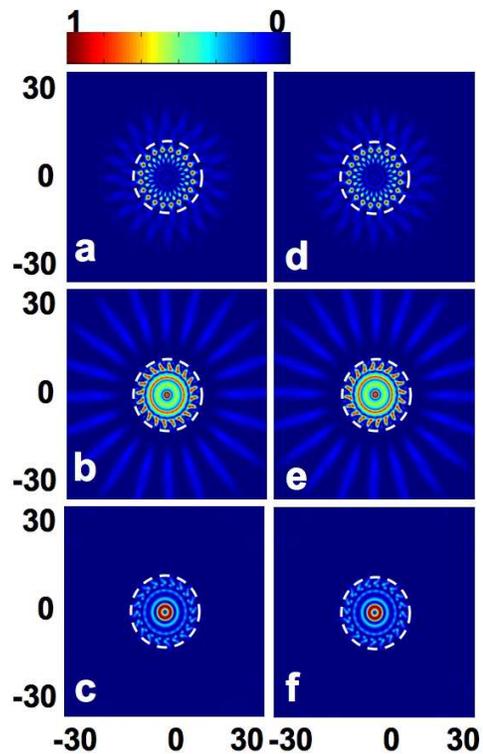}
\caption{Density distributions of a quantized vortex in a
corral-like potential at different times. The left column is for a
vortex rotating counterclockwise at (a) $t=2.5$, (b) $t=3.8$, (c)
$t=6.3$. The right column is for a vortex rotating clockwise at (d)
$t=2.5$, (e) $t=3.8$, (f) $t=6.3$. The coordinates are in units of
$L_{0}$. The dashed circles mark the position of the circular
corrals.}
\end{figure}

In addition to the sunflower-like structure because of discrete
characteristic of the corral-like potential, one more interesting feature
has emerged due to the presence of the vortex. At $t=2.5T_{0}$, as shown in
Fig. 4a and Fig. 4d, the two different vortices have almost identical
density distribution even though their sunflower-like structure indicates
that the two BECs have already felt the corral-potential. The rotational
difference between the two BECs shows up only when the \textquotedblleft
main peak\textquotedblright\ of the condensate hits the corral-potential, as
shown in Figs. 4b-c and Figs. 4e-f. This is due to the unique property of
the velocity field for a quantized vortex, where the center of the vortex
has larger velocity. In Figs. 4c and 4f, a series of flyer-like peaks
distributed along the circumference of the outmost circle distinguish
obviously two vortices with different rotational directions. This feature
may be applied to detect the rotational direction of a vortex.

It is still an open problem to experimentally investigate the behavior of a
quantized vortex in electronic quantum corrals. For an atomic condensate,
all the rich dynamics of a vortex shown above can be readily observed
experimentally as a macroscopically quantized vortex can be generated
experimentally in a BEC with mature technologies\cite{vortex}

\begin{figure}[tbp]
\includegraphics[width=0.85\linewidth,angle=0]{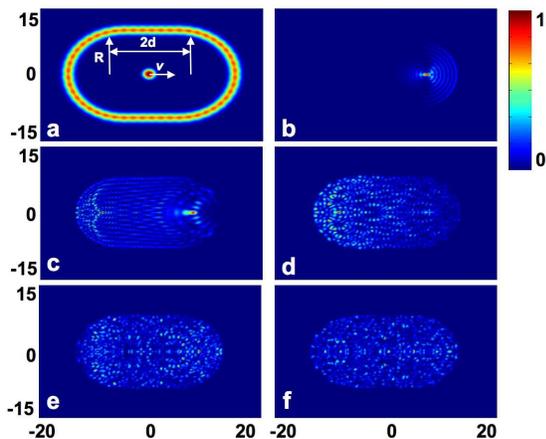}
\caption{Density distributions of a condensate in a
stadium-shaped potential at different times: (a) $t=0$, (b) $t=0.5T$, (c) $%
t=T$, (d) $t=1.5T$, (e) $t=3T$, (f) $t=6T$. $T=2(R+d)/v$ is the time needed
for a classical particle to propagate along the stadium axis. The
coordinates are in units of $L_{0}$. In Fig. (a), the stadium-shaped
confinement potential is also shown.}
\end{figure}

\section{quantum chaos for a BEC in corral-like potential}

We now turn to consider the dynamics of a BEC confined in stadium-shaped
quantum corrals as shown in Fig. 5a. Stadium billiard is a typical system to
study both classical and quantum chaos\cite{stadium}. The energy-level
distribution was proposed in Ref. \cite{Niu} to reveal the quantum chaos for
a BEC. Most recently, the scars in the steady-state density profiles of
parametrically driven condensates was proposed to study the quantum chaos
\cite{Katz}. Our work here focuses on the dynamic manifestation of quantum
chaos with a BEC.

The stadium potential can be described by $V_{oc}$ with $\left\{
x_{i},y_{i}\right\} $ uniformly distributing along the circumference of the
stadium as shown in Fig. 5a. In our numerical computation, the parameters
that we choose for the stadium-shaped quantum corrals are $M=50$, $\sigma =1$%
, $\gamma =500$, $R=10$, and $d=30\pi /19$. The groundstate wave function is
calculated for the condensate confined in the harmonic trap with $\omega
_{\perp }=87.5\pi $ and the stadium-shaped quantum corrals. This condensate
is then given an initial velocity of $v$. After switching off the harmonic
trap, we study numerically the evolution of this moving condensate in the
stadium-shaped quantum corrals. For $v=2.7$ $\mathrm{mm/s}$, Figs. 5a-5f
show a series of snapshots for the condensate evolution. The regular density
distribution is destroyed after a small number of reflections. This is in
stark contrast with the dynamics of a moving BEC in circle-shaped quantum
corrals, which stays regular even after long-time evolution as shown in Fig.
6.


This quantum chaotic behavior was firstly revealed in Ref.\cite{stadium} for
a free particle in a continuous stadium billiard. In this work, Tomsovic and
Heller showed that the chaotic wave dynamics shown in Fig. 5 can be regarded
as the supposition of millions of classical trajectories. The random looking
in Fig. 5 is just a reflection of the chaotic classical trajectories. In
contrast to a free particle, there is interatomic interaction in a BEC and
the corral potential is discrete. As a result, one might expect with slight
hesitation that a BEC in stadium-shaped quantum corrals exhibits typical
quantum chaotic wave dynamics. Our numerical calculations above show that
the hesitation can be cast aside. In fact, no obvious difference is found,
compared to the situation of noninteracting BEC. We have also considered the
situation with increasing coupling constant. It is found that increasing the
coupling constant has the effect of enhancing the dynamical quantum chaotic
behavior.

\begin{figure}[tbp]
\includegraphics[width=0.85\linewidth,angle=0]{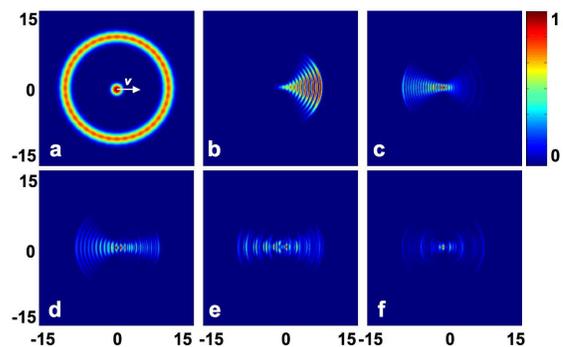}
\caption{Density distributions of a condensate in a
circle-shaped potential at different times: (a) $t=0$, (b) $t=0.5T$, (c) $%
t=T $, (d) $t=1.5T$, (e) $t=3T$, (f) $t=6T$. Here $T=2R/v$. The coordinates
are in units of $L_{0}$. In Fig. (a), the circle-shaped confinement
potential is also shown.}
\end{figure}

As mentioned at the beginning, it has remained elusive to visualize
experimentally the quantum chaotic behavior in the electronic quantum
corrals despite a great deal of effort\cite{Heller_nature}. The main reason
is that the quantum corrals are too \textquotedblleft
leaky\textquotedblright\ for electrons. Our above numerical studies show
that the BEC system provides a very good chance to experimentally visualize
the quantum chaos: The stadium-shaped corrals can be built with the scheme
shown in Fig. 1. The BEC can gain a velocity with the method used for the
experimental studies of quantum reflection\cite{quantumref}. Considering the
length unit is $5$ $\mathrm{\mu m}$ in Figs. 2-6, a imaging resolution below
$5$ $\mathrm{\mu m}$ is necessary to reveal the structure in these figures.
Fortunately, high resolution absorption imaging below $4$ $\mathrm{\mu m}$
has been achieved \cite{Cheng}
The chaotic behavior has been studied experimentally with nondegenerate
ultracold atomic gases\cite{Raizen}; quantum chaos has been visualized with
classical wave systems, such as sound and miscrowave\cite{Stockmann}. An
eventual implementation of our scheme will be an experimental realization of
quantum chaos with coherent matter wave, marking an important step forward
in the experimental studies of chaos.

Besides visualizing the known quantum chaotic behavior, we can try to
explore more on quantum chaos with BECs in a stadium-shaped corrals. For
example, we can study the thermal effect on quantum chaos by placing
non-condensed cold atoms and partially condensed cold atoms into the
corrals. This is impossible for the electronic quantum corrals. One can also
study how the interference between two BECs is affected by quantum chaos. We
have shown in Fig. 7 the evolution of two initially coherently separated
condensates\cite{doubleBECs}. The significance of this study lies in that a
classical particle can not be regarded as two coherently separated
particles. Figs. 7(a1)-7(a6) give the evolution of two coherently separated
Gaussian wavepackets with the same condition in Fig. 5. The initial width
and distance are $\sqrt{2/5}$ and $4$. For short-time evolution, the
interference between two condensates is seen in Fig. 7(a2). After long-time
evolution, the interference is destroyed by quantum chaos as seen in Figs.
7(a4)-7(a6). As a comparison, we have also computed the evolution of these
two BECs in circle-shaped corrals. The results are shown in Fig. 7(b1-b6),
where the interference between two condensates is preserved partly even for
long-time evolution. In the inset of Fig. 7(b5), the fork-like structure
implies the information of two initially coherently separated condensates.

\begin{figure}[tbp]
\includegraphics[width=1.3\linewidth,angle=0]{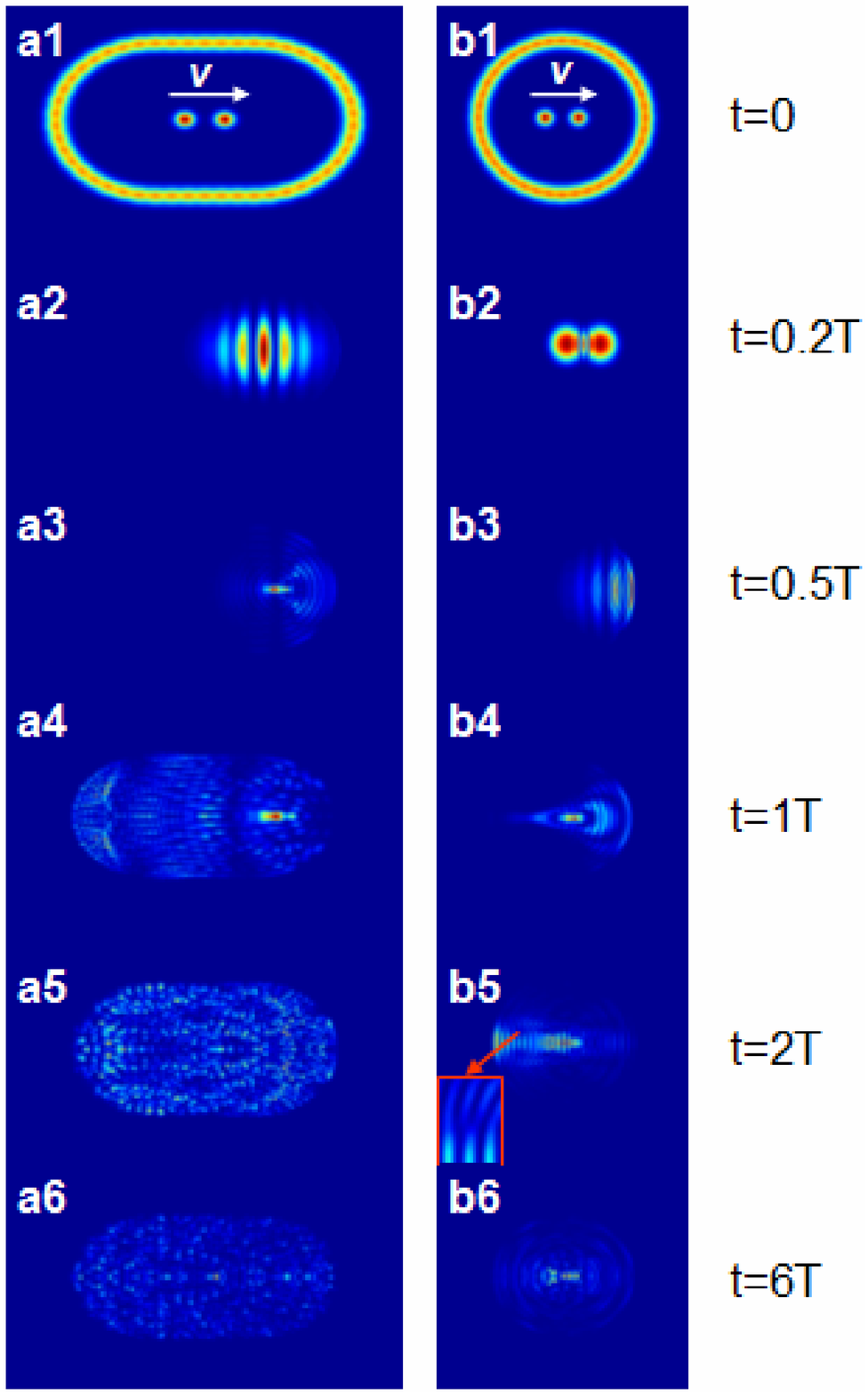}
\caption{The left and right columns give the evolution of two
initially coherently separated condensates in a stadium-shaped and
circle-shaped potentials, respectively. Each figure in the left and
right columns is $30\times 40$ and $30\times 30$ with coordinate
unit $L_{0}$. Here the time unit is $T=2\left( R+d\right) /v$ and
$T=2R/v$ for the left and right columns.}
\end{figure}

\section{summary and discussion}

In summary, we have studied an atomic analogy to electronic quantum corrals.
A scheme is proposed to study the dynamic evolution of a Bose-Einstein
condensate confined in a corral-like potential. In particular, with these
atomic corrals, it is now promising to study experimentally the dynamic
quantum chaotic behavior \cite{stadium}. The atomic quantum corrals proposed
here can also be applied to study other quantum gases, such as molecular
Bose-Einstein condensates \cite{molecularBEC}, degenerate Fermi gases \cite%
{degenerageFerm}, ultracold Fermi gases in the unitarity limit \cite%
{Unitarity}, and BCS (Bardeen-Cooper-Schrieffer) superfluid \cite%
{BCSsuperfluid} in the corral-like potential. The different descriptions of
the order parameter, quantum statistics and evolution equation \cite%
{Fermi-rev} mean that there should be very rich phenomena for discovery.

\begin{acknowledgments}
We acknowledge useful discussions with Baolong L\"{u}, Cheng Chin, and Jin
Wang. This work was supported by NSFC under Grant Nos. 10875165, 10825417,
10634060, and NBRPC 2006CB921406.
\end{acknowledgments}


\begin{thebibliography}{99}
\bibitem{eleccroo} M. F. Crommie, C. P. Lutz, and D. M. Eigler, Nature
(London) \textbf{363}, 524 (1993); M. F. Crommie, C. P. Lutz, and D. M.
Eigler, Science \textbf{262}, 218 (1993).

\bibitem{Heller_rmp} G. A. Fiete and E. J. Heller, Rev. Mod. Phys. \textbf{75%
}, 933 (2003).

\bibitem{theo-opt} G. C. des Francs \textit{et al}., Phys. Rev. Lett.
\textbf{86}, 4950 (2001).

\bibitem{exp-opt} C. Chicanne \textit{et al}., Phys. Rev. Lett. \textbf{88},
097402 (2002).

\bibitem{Ketterle-Rev} Nature Insight: Ultracold Matter [Nature (London)
\textbf{416}, 205 (2002)]; J. P. Yin, Phys. Rep. \textbf{430}, 1 (2006).

\bibitem{Rev} F. Dalfovo, S. Giorgini, L. P. Pitaevskii, and S. Stringari,
Rev. Mod. Phys. \textbf{71}, 463 (1999); A. J. Leggett, Rev. Mod. Phys.
\textbf{73}, 307 (2001); C. J. Pethick and H. Smith, \textit{Bose-Einstein
Condensation in Dilute Gases} (Cambridge University, Cambridge, 2002); L. P.
Pitaevskii and S. Stringari, \textit{Bose-Einstein condensation} (Clarendon,
Oxford, 2003).

\bibitem{Heller_nature} E. J. Heller, M. F. Crommie, C. P. Lutz, and D. M.
Eigler, Nature \textbf{369}, 464 (1994); M. F. Crommie, C. P. Lutz, D. M.
Eigler, and E. J. Heller, Physica D \textbf{83}, 98 (1995); M. F. Crommie,
C. P. Lutz, D. M. Eigler, and E. J. Heller, Surf. Sci. \textbf{361}, 864
(1996).

\bibitem{stadium} S. Tomsovic and E. J. Heller, Phys. Rev. Lett. \textbf{67}%
, 664 (1991); Phys. Rev. E \textbf{47}, 282 (1993).

\bibitem{Tung} S. Tung, V. Schweikhard and E. A. Cornell, Phys. Rev. Lett.
\textbf{97}, 240402 (2006).

\bibitem{random} D. Clement \textit{et al}., New J. of Phys. \textbf{8}, 165
(2006).

\bibitem{Lv} K. Henderson \textit{et al}., Preprint arXiv:0902.2171 (2009);
A. Itah \textit{et al}., Preprint arXiv:0903.3282 (2009); V. Boyer \textit{%
et al}., Phys. Rev. A \textbf{73}, 031402(R) (2006).

\bibitem{vortex} A. L. Fetter, Preprint arXiv:0801.2952 (2008); M. R.
Matthews \textit{et al}., Phys. Rev. Lett. \textbf{83}, 2498 (1999); K. W.
Madison, F. Chevy, W. Wohlleben, and J. Dalibard, Phys. Rev. Lett. \textbf{84%
}, 806 (2000); J. R. Abo-Shaeer, C. Raman, J. M. Vogels, and W. Ketterle,
Science \textbf{292}, 476 (2001). 

\bibitem{Niu} C. W. Zhang, J. Liu, M. G. Raizen, and Q. Niu, Phys. Rev.
Lett. \textbf{93}, 074101 (2004).

\bibitem{Katz} N. Katz and O. Agam, Preprint arXiv: 0903.0968 (2009).

\bibitem{quantumref} T. A. Pasquini \textit{et al}., Phys. Rev. Lett.
\textbf{93}, 223201 (2004).

\bibitem{Cheng} N. Gemelke \textit{et al}., Preprint arXiv:0904.1532 (2009);
private communication with Dr. Cheng Chin about the improved imaging
resolution of about $1.5$ $\mathrm{\mu m}$.

\bibitem{Raizen} V. Milner, J. L. Hanssen, W. Campbell, and M. G. Raizen,
Phys. Rev. Lett. \textbf{86}, 1514 (2001); D. A. Steck, W. H. Oskay, and M.
G. Raizen, Science. \textbf{293}, 274 (2001); N. Friedman \textit{et al}.,
Phys. Rev. Lett. \textbf{86}, 1518 (2001).

\bibitem{Stockmann} H. J. St\"ockmann, \textit{Quantum Chaos: an Introduction%
} (Cambridge, Cambridge, 1999).

\bibitem{doubleBECs} M. R. Andrews \textit{et al}., Science \textbf{275},
637 (1997).

\bibitem{molecularBEC} S. Jochim \textit{et al}., Science \textbf{302}, 2101
(2003); M. Greiner, C.~A. Regal, and D.~S. Jin, Nature \textbf{426}, 537
(2003); M. W. Zwierlein \textit{et al}., Phys. Rev. Lett. \textbf{91},
250401 (2003).

\bibitem{degenerageFerm} B. D. Marco, and D. S. Jin, Science \textbf{285},
1703 (1999).

\bibitem{Unitarity} K. M. O'Hara \textit{et al}., Science \textbf{298}, 2179
(2002).

\bibitem{BCSsuperfluid} C. A. Regal, M. Greiner, and D. S. Jin, Phys. Rev.
Lett. \textbf{92}, 040403 (2004); M. W. Zwierlein \textit{et al}., Phys.
Rev. Lett. \textbf{92}, 120403 (2004); M. Bartenstein \textit{et al}., Phys.
Rev. Lett. \textbf{92}, 120401 (2004); J. Kinast \textit{et al}., Phys. Rev.
Lett. \textbf{92}, 150402 (2004).

\bibitem{Fermi-rev} S. Giorgini, L. P. Pitaevskii, and S. Stringari, Rev.
Mod. Phys. \textbf{80}, 1215 (2008).
\end{thebibliography}
\end{document}